\documentclass[twoside,12pt]{article}
\newcommand{\beq}{\begin{equation}}
\newcommand{\eeq}{\end{equation}}
\usepackage{epsfig}
\newcommand\lsim{\mathrel{\rlap{\lower4pt\hbox{\hskip1pt$\sim$}}
    \raise1pt\hbox{$<$}}}
\newcommand\gsim{\mathrel{\rlap{\lower4pt\hbox{\hskip1pt$\sim$}}
    \raise1pt\hbox{$>$}}}

\newcommand\lesssim{\mathrel{\rlap{\lower4pt\hbox{\hskip1pt$\sim$}}
    \raise1pt\hbox{$<$}}}
\newcommand\gtrsim{\mathrel{\rlap{\lower4pt\hbox{\hskip1pt$\sim$}}
    \raise1pt\hbox{$>$}}}

\begin{document}

\title{Dust-Filled Axially Symmetric Universes \\ with
Cosmological Constant}
\author{{\bf Paulo Aguiar}\thanks{Email address: paguiar@cosmo.cii.fc.ul.pt}
and {\bf Paulo Crawford}\thanks{Email address:
crawford@cosmo.cii.fc.ul.pt}\vspace*{0.5cm} \\
{\em Centro de F\'{\i}sica Nuclear e Departamento de F\'{\i}sica} \\
{\em da Faculdade de Ci\^enicas da Universidade de Lisboa} \\
{\em Av. Prof. Gama Pinto, 2; 1649-003 Lisboa -- Portugal}}
\date{}
\maketitle

\begin{abstract}
Following the recent recognition of a positive value for the
vacuum energy density and the realization that a simple
Kantowski-Sachs model might fit the classical tests of cosmology,
we study the qualitative behavior of three anisotropic and
homogeneous models: Kantowski-Sachs, Bianchi type-I and Bianchi
type-III universes, with dust and a cosmological constant, in
order to find out which are physically permitted. We find that
these models undergo isotropization up to the point that the
observations will not be able to distinguish between them and the
standard model, except for the Kantowski-Sachs model
$(\Omega_{k_{0}}<0)$ and for the Bianchi type-III
$(\Omega_{k_{0}}>0)$ with $\Omega_{\Lambda_{0}}$ smaller than some
critical value $\Omega_{\Lambda_{M}}$. Even if one imposes that
the Universe should be nearly isotropic since the last scattering
epoch ($z\approx 1000$), meaning that the Universe should have
approximately the same Hubble parameter in all directions
(considering the COBE 4-Year data), there is still a large range
for the matter density parameter compatible with Kantowsky-Sachs
and Bianchi type-III if $|\Omega_0+\Omega_{\Lambda_0}-1|\leq
\delta$, for a very small $\delta$ . The Bianchi type-I model
becomes exactly isotropic owing to our restrictions and we have
$\Omega_0+\Omega_{\Lambda_0}=1$ in this case. Of course, all these
models approach locally an exponential expanding state provided
the cosmological constant $\Omega_\Lambda>\Omega_{\Lambda_{M}}$.
\end{abstract}


\section{Introduction}

Over the last five years, the issue of whether or not there is a
nonzero value for the vacuum energy density, or cosmological
constant, has been raised quite often. Indeed, the possibility of
a nonzero cosmological constant $\Lambda$ has been entertained
several times in the past for theoretical  and observational
reasons (the early history of $\Lambda$ as a parameter in General
Relativity has been reviewed by \cite{North}, \cite{Petrosian},
and \cite{Gunn}). Recent supernovae results \cite{Perl},
\cite{Riess} strongly support a positive and possibly quite large
cosmological constant. Even taking the Hubble constant to be in
the range 60-75 km/s/Mpc it is possible to show \cite{Roos} that
the standard model of flat space with vanishing cosmological
constant is ruled out. In a very nice review \cite{Carroll} it is
argued that postulating an $\Omega_\Lambda$-dominated model seems
to solve a lot of problems at once. And again, in a quite recent
review on the physics and cosmology of the cosmological constant,
it is added that ``recent years have provided the best evidence
yet that this elusive quantity does play an important dynamical
role in the universe" \cite{Carroll2}.

On the other hand,
if the classical tests of cosmology are applied to a simple
Kantowski-Sachs metric and the results compared with those obtained for the
standard model, the observations will not be able to distinguish
between these models if the Hubble parameters along the
orthogonal directions are assumed to be approximately equal
\cite{Henriques}.
Indeed, as \cite{Collins} points out, the number
of cosmological solutions which demonstrate exact isotropy well
after the big bang origin of the Universe is a small fraction of
the set of allowable solutions of the cosmological equations.
It is therefore prudent to take seriously the possibility that
the Universe is expanding anisotropically. Note also that some shear free
anisotropic models display a FLRW-like behavior, as it is shown in
\cite{Crawford}.

\section{ The global behavior of the $\Lambda \ne 0$ models}

Taking all this into consideration, we discuss the behavior of
some homogeneous but anisotropic models with axial symmetry,
filled with a pressureless
perfect fluid (dust) and a non vanishing cosmological constant.
For this, we will restrict our study to the the metric forms
\begin{equation}
ds^2=-c^2dt^2+a^2(t)dr^2+b^2(t)\left(d\theta^2+
f^2_k(\theta)d\phi^2\right),
\label{metric}
\end{equation}
with the two scale factors $a(t)$ and $b(t)$; $k$ is the curvature
index of the 2-dimensional surface $d\theta^2+f^2_k(\theta)d\phi^2$
and can take the values $+1, 0, -1$ implying $f_k(\theta)$
equal $\sin(\theta)$,
$\theta$, $\sinh(\theta)$, respectively, giving the following three
different metrics: Kantowski-Sachs, Bianchi type-I, and Bianchi type-III
\cite{Burd}, \cite{Byland}.

Einstein equations for the metric (\ref{metric}), for which
the matter content is in the form of a perfect fluid and
a cosmological term, $\Lambda$, are then as follows
\cite{Burd}, \cite{Byland}:
\begin{equation}
2~\frac{\dot a}{a}~\frac{\dot b}{b}+\frac{\dot b^2}{b^2}+\frac{kc^2}{b^2}
=8\pi G\rho+\Lambda c^2,
\label{EE1a}
\end{equation}
\begin{equation}
2~\frac{\ddot b}{b}+\frac{\dot b^2}{b^2}+\frac{kc^2}{b^2}
=-8\pi G \frac{p}{c}+\Lambda c^2,
\label{EE1b}
\end{equation}
\begin{equation}
\frac{\ddot a}{a}+\frac{\ddot b}{b}+\frac{\dot a}{a}~\frac{\dot b}{b}
=-8\pi G \frac{p}{c}+\Lambda c^2,
\label{EE1c}
\end{equation}
where $\rho$ is the matter density and $p$ is the (isotropic)
pressure of the fluid. Here $G$ is the Newton's gravitational
constant and $c$ is the speed of light. If we consider a
vanishing pressure $(p=0)$, which is compatible with the present
conditions for the Universe, the last two equations take the form
\begin{equation}
2~\frac{\ddot b}{b}+\frac{\dot b^2}{b^2}+\frac{kc^2}{b^2}
=\Lambda c^2,
\label{EE2b}
\end{equation}
\begin{equation}
\frac{\ddot a}{a}+\frac{\ddot b}{b}+\frac{\dot a}{a}~\frac{\dot b}{b}
=\Lambda c^2,
\label{EE2c}
\end{equation}
and equation (\ref{EE2b}) can easily be integrated to give
\begin{equation}
\frac{\dot b^2}{b^2}=\frac{M_1}{b^3}-\frac{kc^2}{b^2}+
\frac{\Lambda}{3}c^2,
\label{eqC1}
\end{equation}
where $M_1$ is a constant of integration.

The Hubble parameters corresponding to the scale factors $a(t)$ and $b(t)$
are defined by

$$H_a\equiv\dot a/a \quad \mbox{and} \quad H_b \equiv  \dot b/b.$$

Using them to introduce the following dimensionless parameters,
in analogy with which it is usually done in the
Friedmann-Lema\^{\i}tre-Robertson-Walker (FLRW) universes,
let us define
\begin{equation}
\frac{M_1}{b^3 H_b^2}\equiv \Omega_M,
\end{equation}
\begin{equation}
-\frac{kc^2}{b^2 H_b^2}\equiv \Omega_k
\end{equation}
and
\begin{equation}
\frac{\Lambda c^2}{3 H_b^2}\equiv \Omega_{\Lambda}.
\end{equation}
The conservation equation (\ref{eqC1}) can now be rewritten as
\begin{equation}
\Omega_M+\Omega_k+\Omega_{\Lambda}=1.
\label{eqFG2}
\end{equation}

Now defining the dimensionless variable $y=b/b_0$ where $b_0=b(t_0)$
is the angular scale factor for the present age of the Universe, and
using equation (\ref{eqFG2}) (taken for $t=t_0$), one may rewrite
equation (\ref{eqC1}) as
\begin{equation}
\dot y=\pm H_{b_{0}}\sqrt{\Omega_{M_{0}}(\frac{1}{y}-1)+
\Omega_{\Lambda_{0}}(y^2-1)+1},
\label{eqy}
\end{equation}
where the density parameters defined previously and $H_b$
with the zero subscript,
denote as before these quantities at the present time $t_0$. In this
way, the number of independent parameters have been reduced.
Substituting equation (\ref{eqC1}) into equation (\ref{EE1a}) gives
\begin{equation}
\dot a=\frac{M_{\rho}-M_1~\frac{a}{b}+\frac{2}{3}\Lambda c^2ab^2}
{2\sqrt{M_1b-kc^2b^2+\frac{\Lambda}{3}c^2b^4}},
\label{dota}
\end{equation}
where $M_{\rho}$ is a constant proportional to the matter in the
Universe,
\begin{equation}
M_{\rho}=8\pi G\rho ab^2.
\end{equation}
Using the procedure above, equation (\ref{dota}) can be rewritten in the form
\begin{equation}
\Omega_{\rho}-\Omega_M+2~\Omega_{\Lambda}=2~\frac{H_a}{H_b},
\label{eqC3}
\end{equation}
where
\begin{equation}
\Omega_{\rho}=\frac{M_{\rho}}{ab^2H_b^2}.
\label{eqC4}
\end{equation}
From equation (\ref{EE1a}) one may define a matter density
parameter. For this, we introduce the notion of mean Hubble
factor $H$ such that $3H=H_a+2H_b$. Also, for these models, the
shear scalar $\sigma$ \cite{Byland} is given by
\begin{equation}
\sigma=\frac{1}{\sqrt{3}}(H_a-H_b)
\end{equation}
Thus, equation (\ref{EE1a}) may be rewritten \cite{Burd} as
\begin{equation}
3H^2+\frac{kc^2}{b^2}=8\pi G\rho+\sigma^2+\Lambda c^2.
\end{equation}
As in FLRW we call critical matter density $\rho_c$ when $k=0$
and $\Lambda=0$
\begin{equation}
\rho_c=\frac{3H^2-\sigma^2}{8\pi G}.
\end{equation}
The matter density is generally defined as $\Omega=\rho/\rho_c$,
then
\begin{equation}
\Omega=\frac{8\pi G \rho}{3H^2-\sigma^2}
\equiv {{8\pi G\rho} \over {2H_aH_b+H_b^2}},
\end{equation}
just like in FLRW models, and such that $\Omega=1$ when $k=0$ and
$\Lambda=0$, and which is related to $\Omega_{\rho}$ by
\begin{equation}
\Omega ={{\Omega _\rho} \over {1+2{{H_a} \over {H_b}}}}.
\label{omega}
\end{equation}
Although $\Omega_M$ is not the matter density parameter, it performs the
same important role.
We emphasize the fact that if for one particular time $H_a=H_b$
and $\Omega_{\Lambda}=1$, then, by equations (\ref{eqFG2}), (\ref{eqC3})
and (\ref{omega}),
$3\Omega=\Omega_{\rho}=\Omega_{M}=-\Omega_{k}$;
and if $ 0<\Omega_{\Lambda}\ll 1$ and $\Omega_M =1$, then,
$-\Omega_k=\Omega_{\Lambda}$ and $\Omega\simeq 1$.

Introducing another dimensionless variable $x=a/a_0$,
equation (\ref{dota}) takes the form
\begin{equation}
\dot x=H_{b_{0}}~\frac{\frac{\Omega_{M_{0}}}{2}(1-\frac{x}{y})
+\Omega_{\Lambda_{0}}(-1+xy^2)+\frac{H_{a_{0}}}{H_{b_{0}}}}
{y\sqrt{\Omega_{M_{0}}(\frac{1}{y}-1)+\Omega_{\Lambda_{0}}
(y^2-1)+1}},
\label{eqx}
\end{equation}
and its number of independent parameters was also reduced, now at
the expense of equation (\ref{eqC3}) taken for the present time $t=t_0$.

Now, we want to find the time dependence of $b(t)$ in a
qualitative way, starting from equation (\ref{eqy}). Since the
model universe will be defined only where $\dot y^2\ge 0$, as was
previously done by \cite{Moles} for FLRW models, the problem is
reduced to find out the zeros of $\dot y$, with $y\neq 0$.

There are two $\Omega_{\Lambda}$ values which characterize two
zones of distinct behavior for the scale factor $b$. Starting with
condition $\dot y=0$ one may obtain
\begin{equation}
\Omega_{\Lambda_{0}}=\frac{(\Omega_{M_{0}}-1)y-\Omega_{M_{0}}}{y^3-y}
\end{equation}
If we consider $\Omega_{\Lambda_{0}}=\Omega_{\Lambda_{0}}(y)$, as
a function of $y$, then
this function presents a relative minimum and a maximum, that we will
denote by $\Omega_{\Lambda_{c}}$ and $\Omega_{\Lambda_{M}}$, respectively.
The relative minimum depends on $\Omega_{M_{0}}$ in the following way
\cite{Moles}:
For $\Omega_{M_{0}}<1/2$
we have
\begin{eqnarray}
\Omega_{\Lambda_{c}} & = & \frac{3\Omega_{M_{0}}}{2}\left\{ \left[
\sqrt{ \frac{(\Omega_{M_{0}}-1)^2}{\Omega_{M_{0}}^2}-1}+
\frac{1-\Omega_{M_{0}}}{\Omega_{M_{0}}} \right]^{1/3}+
\right. \nonumber
\\ & &
\left.
\frac{1}{\left[ \sqrt{(\Omega_{M_{0}}-1)^2/\Omega_{M_{0}}^2-1
}+(1-\Omega_{M_{0}})/\Omega_{M_{0}} \right]^{1/3}} \right\}-
(\Omega_{M_{0}}-1),
\end{eqnarray}
for $\Omega_{M_{0}}\geq 1/2$ the expression is
\begin{equation}
\Omega_{\Lambda_{c}}=-3\Omega_{M_{0}}\cos\left(
\frac{\theta+2\pi}{3}\right) -(\Omega_{M_{0}}-1).
\end{equation}
The relative maximum is done by
\begin{equation}
\Omega_{\Lambda_{M}}=-3\Omega_{M_{0}}\cos \left(
\frac{\theta+4\pi}{3}\right) -(\Omega_{M_{0}}-1),
\end{equation}
where $\theta=\cos^{-1}\left( \frac{\Omega_{M_{0}}-1}
{\Omega_{M_{0}}} \right)$. These expressions are limiting zones
of the ($\Omega_{\Lambda_{0}}, \Omega_{M_{0}}$) plane, where
$\dot y=0$ has three or one solutions (for details see
\cite{Moles}). The $\Omega_{\Lambda_{M}}$ expression is also
defined for $\Omega_{M_{0}}>1/2$, but it has the meaning of a
maximum only for $\Omega_{M_{0}}>1$. The $\Omega_{\Lambda_{0}}$
less or equal to $\Omega_{\Lambda_{M}}$ imposes the recollapse of
scale factor $b$, while greater values produces inflexional
behaviors for $b$. The $\Omega_{\Lambda_{0}}$ values greater or
equal to $\Omega_{\Lambda_{c}}$ are physically ``forbidden"
because they don't reproduce the present Universe (see
\cite{Moles}). Obviously, $\Omega_{\Lambda_{M}} <
\Omega_{\Lambda_{c}}$ always.

Although we are considering anisotropic models, the equation
(\ref{eqy}) for $\dot y$ as a function of $\Omega_{M_{0}}$ is
mathematically the same as equation (2) obtained by \cite{Moles}
for the homogeneous and isotropic FLRW models. From  equations
(\ref{eqy}) and (\ref{eqx}) one obtains the differential equation
\begin{equation}
\frac{dx}{dy}=\frac{\frac{\Omega_{M_{0}}}{2}(1-\frac{x}{y})
+\Omega_{\Lambda_{0}}(-1+xy^2)+\frac{H_{a_{0}}}{H_{b_{0}}}}
{\Omega_{M_{0}}(1-y)+\Omega_{\Lambda_{0}}(y^3-y)+y}.
\label{dif}
\end{equation}

This equation automatically complies with the two conservation equations
(\ref{eqFG2}) and (\ref{eqC3}) evaluated at $t_0$.
There are some particular values of the parameters ($\Omega_{M_{0}},
\Omega_{\Lambda_{0}}$)
for which this equation has exact solutions. However, for the
majority of the values of the parameters, the solution has only been
obtained by numerical integration.

We may admit that
at a certain moment of time, which we can take as the present time
$t_0$, the Hubble parameters along the orthogonal directions may be
assumed to be approximately equal, $H_a\simeq H_b$, even though we
started with an anisotropic geometry.
This hypothesis has been considered in \cite{Henriques}
for the case of a Kantowski-Sachs (KS) model. From this study was
derived the conclusion that the classical tests of cosmology are
not at present sufficiently accurate to distinguish between a
FLRW model and the KS defined in that paper, with
$(H_{a_{0}}\simeq H_{b_{0}})$, except for small values of $b_0$.

Taking $H_{a_{0}}=H_{b_{0}}$ for simplicity, one can then
integrate equation (\ref{dif}) and find three different
solutions, one for each $k$ value. Figure 1 and Figure 2 shows the
three kinds of behaviors as a result of integration.
\begin{figure}[h!]
\centerline{\epsfig{file=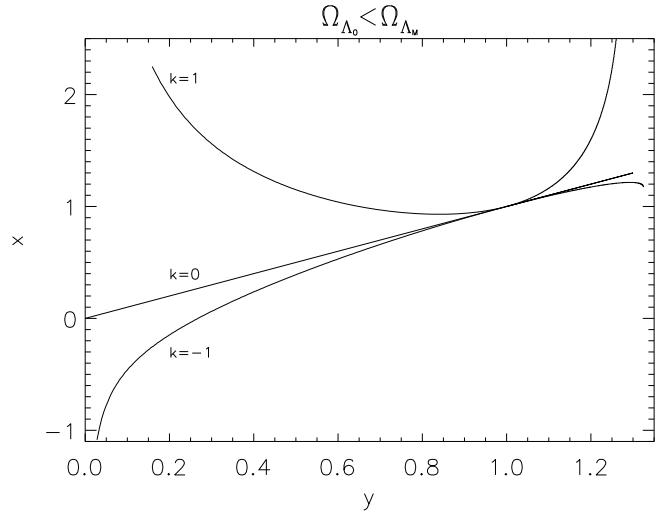,width=10.cm, angle=0}}
\caption{Scale factors relation, that is, the $x~y$ dependence for
the three models Kantowski-Sachs ($k=1$), Bianchi type-I ($k=0$)
and Bianchi type-III ($k=-1$). We show the behaviour of $x(y)$
when $\Omega_{\Lambda_{0}}<\Omega_{\Lambda_{M}}$. Concretely we
have for Kantowski-Sachs $\Omega_{M_{0}}=9$ and
$\Omega_{\Lambda_{0}}=1.5$; for Bainchi type-I $\Omega_{M_{0}}=2$
and $\Omega_{\Lambda_{0}}=-1$; for Bainchi type-III
$\Omega_{M_{0}}=1$ and $\Omega_{\Lambda_{0}}=-1$. \label{fig1}}
\end{figure}
\begin{figure}[h!]
\centerline{\epsfig{file=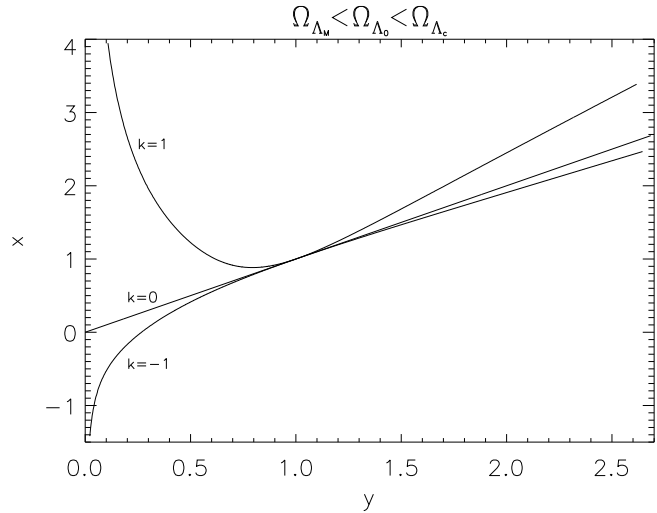,width=10.cm, angle=0}}
\caption{Scale factors relation, that is, the $x~y$ dependence for
the three models Kantowski-Sachs ($k=1$), Bianchi type-I ($k=0$)
and Bianchi type-III ($k=-1$). We show the behaviour of $x(y)$
when
$\Omega_{\Lambda_{M}}<\Omega_{\Lambda_{0}}<\Omega_{\Lambda_{c}}$.
The particular values for the plotting are, for Kantowski-Sachs
$\Omega_{M_{0}}=2$ and $\Omega_{\Lambda_{0}}=1.5$; for Bainchi
type-I $\Omega_{M_{0}}=0.5$ and $\Omega_{\Lambda_{0}}=0.5$; for
Bianch type-III $\Omega_{M_{0}}=0.2$ and
$\Omega_{\Lambda_{0}}=0.4$. \label{fig2}}
\end{figure}

The behavior for the Kantowski-Sachs and Bianchi type-III cases
depends on the $\Omega_{\Lambda_{0}}$ value. If
$\Omega_{\Lambda_{0}}\leq \Omega_{\Lambda_{M}}$, there will be a
maximum value for $y, (y_m),$ and since then $\dot y(y_m)=0$, the
slope of the curve $x=x(y)$ will be infinite at that point.
Specifically, we have $\dot x(y_m)=+\infty$ for Kantowski-Sachs,
and $\dot x(y_m)=-\infty$ for Bianchi type-III, even though
$x(y_m)$ is finite. When
$\Omega_{\Lambda_{M}}<\Omega_{\Lambda_{0}}<
\Omega_{\Lambda_{c}}$, after $x=y=1$ is reached we find an almost
linear relation between the two scale factors $x$ and $y$ for the
two models. While for Bianchi type-I model we have $x=y$ for the
present restrictions. So, we see that for the KS model, the scale
factor $a(t)$ starts from infinity if $b(t)$ starts from zero.
For the Bianchi type-I model, the scale factors are always
proportional or even equal. In this situation we don't have an
anisotropic model; in fact, we can easily prove that this model
is isotropic by a properly reparametrization of the coordinates.
For the Bianchi type-III model, the scale factor $b(t)$ never
starts from zero, but has an initial value different from zero
when $a$ is null. The following plot shows the zones in the
2-dimensional parameter space ($\Omega_{M_0},
\Omega_{\Lambda_0}$) where each model is allowed (Figure 3).
\begin{figure}[h!]
\centerline{\epsfig{file=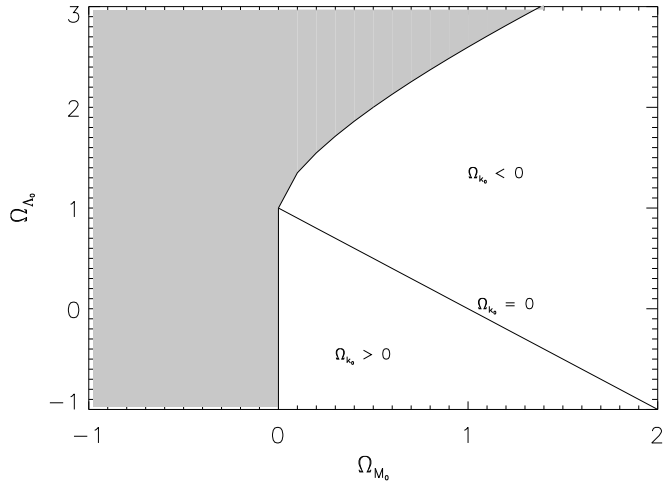,width=10.cm, angle=0}}
\caption{The Kantowski-Sachs model corresponds to the region above
the straight line ($\Omega_{k_0}<0$); the Bianchi type-III model
corresponds the region below the straight line;
($\Omega_{k_0}>0$); the straight line represents the set of region
for the Bianchi type-I model ($\Omega_{k_0}=0$). \label{fig3}}
\end{figure}

Taking into account the analysis given in \cite{Moles}, we may
easily derive the qualitative behavior of $y(t)$, since our
equation (\ref{eqy}) is mathematically equivalent to his equation
(3). Now, going back to Figure 1 and Figure 2, one can then
determine the $x(t)$. The plotting below summarizes the several
possibilities for the three models: Kantowski-Sachs, Bianchi
type-I and Bianchi type-III models, respectively.
\begin{figure}[h!]
\centerline{\epsfig{file=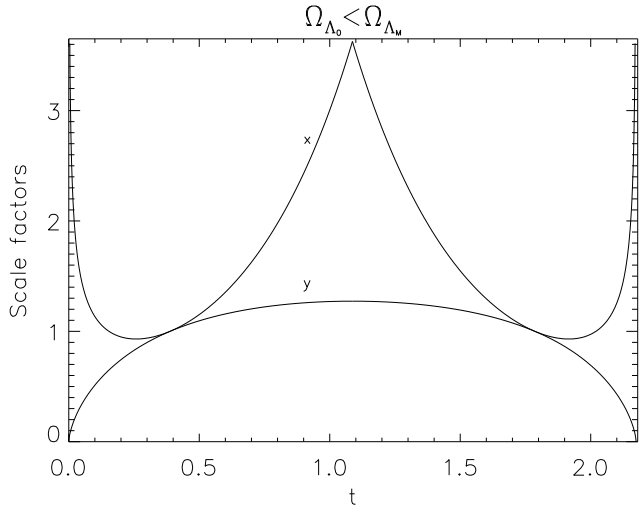,width=10.cm, angle=0}}
\caption{The scale factors $x$ and $y$ for the Kantowski-Sachs
model $(\Omega_{k_{0}}<0)$ when
$\Omega_{\Lambda_{0}}<\Omega_{\Lambda_{M}}$. For the plotting we
put $\Omega_{M_{0}}=9$ and $\Omega_{\Lambda_{0}}=1.5$.
\label{fig4}}
\end{figure}
\begin{figure}[h!]
\centerline{\epsfig{file=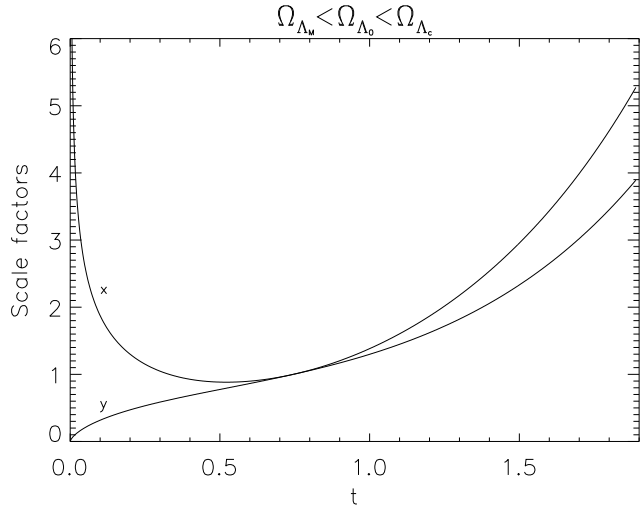,width=10.cm, angle=0}}
\caption{The scale factors $x$ and $y$ for the Kantowski-Sachs
model $(\Omega_{k_{0}}<0)$ when
$\Omega_{\Lambda_{M}}<\Omega_{\Lambda_{0}} <\Omega_{\Lambda_{c}}$.
For the plotting we put $\Omega_{M_{0}}=2$ and
$\Omega_{\Lambda_{0}}=2$. \label{fig5}}
\end{figure}
\begin{figure}[h!]
\centerline{\epsfig{file=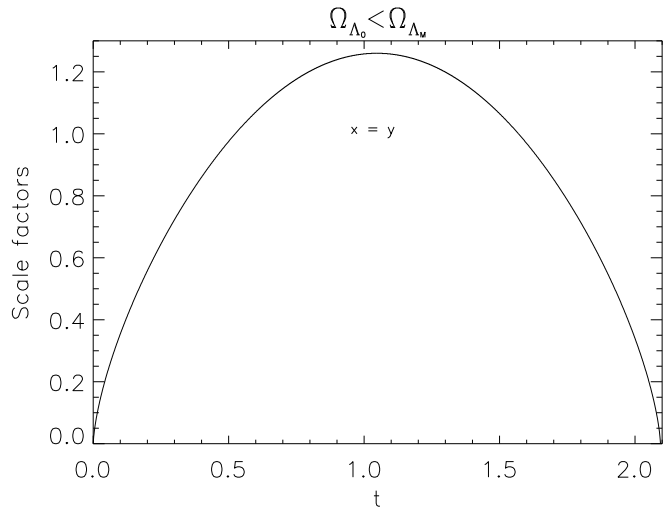,width=10.cm, angle=0}}
\caption{The scale factors $x$ and $y$ for the Bianchi type-I
model $(\Omega_{k_{0}}=0)$ when $\Omega_{\Lambda_{0}}<0$. For the
plotting we put $\Omega_{M_{0}}=2$ and $\Omega_{\Lambda_{0}}=-1$.
\label{fig6}}
\end{figure}
\begin{figure}[h!]
\centerline{\epsfig{file=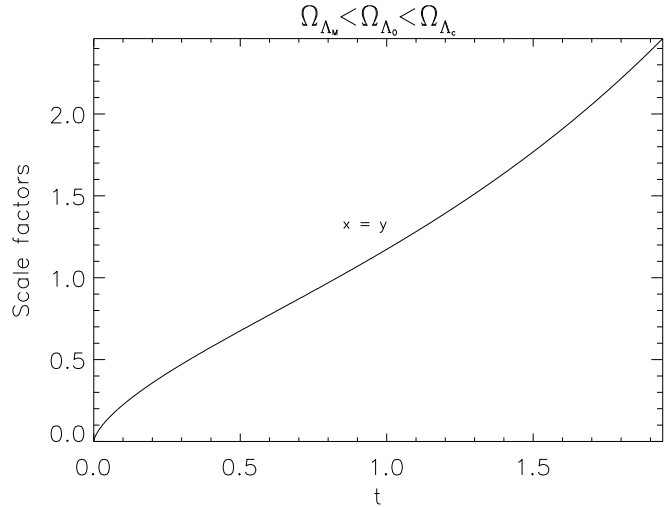,width=10.cm, angle=0}}
\caption{The scale factors $x$ and $y$ for Bianchi type-I model
$(\Omega_{k_{0}}=0)$ when $\Omega_{\Lambda_{0}}\geq 0$. For the
plotting we put $\Omega_{M_{0}}=0.5$ and
$\Omega_{\Lambda_{0}}=0.5$. \label{fig7}}
\end{figure}
\begin{figure}[h!]
\centerline{\epsfig{file=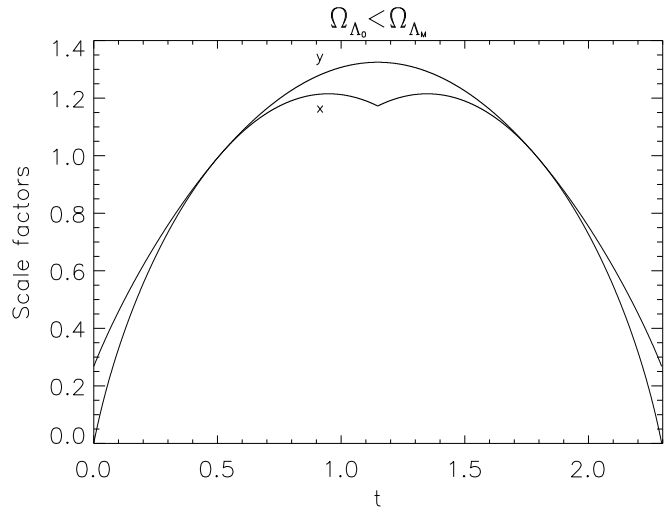,width=10.cm, angle=0}}
\caption{The scale factors $x$ and $y$ for the Bianchi type-III
model $(\Omega_{k_{0}}>0)$ when
$\Omega_{\Lambda_{0}}<\Omega_{\Lambda_{M}}$. For the plotting we
put $\Omega_{M_{0}}=1$ and $\Omega_{\Lambda_{0}}=-1$.
\label{fig8}}
\end{figure}
\begin{figure}[h!]
\centerline{\epsfig{file=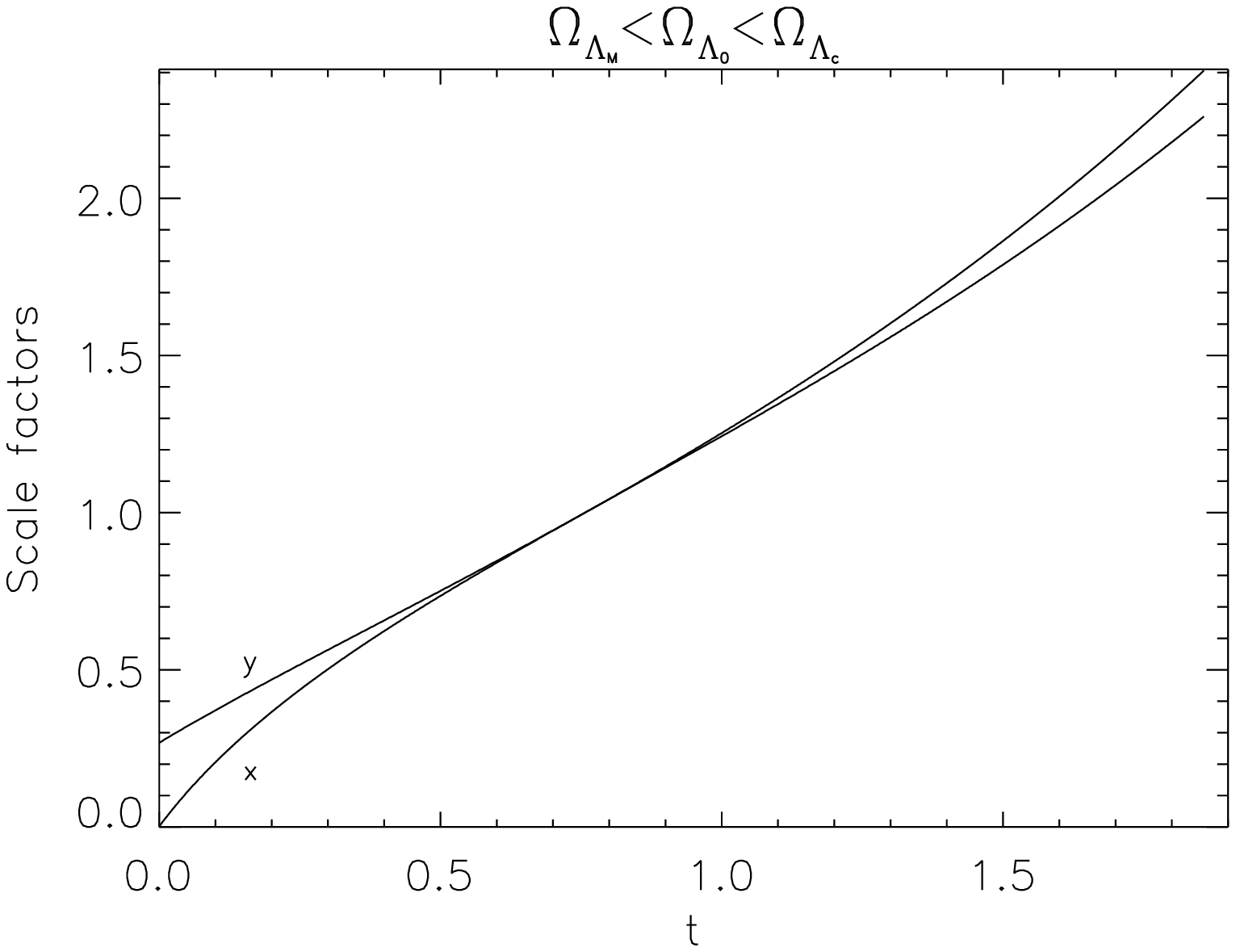,width=10.cm, angle=0}}
\caption{The scale factors $x$ and $y$ for the Bianchi type-III
model $(\Omega_{k_{0}}>0)$ when
$\Omega_{\Lambda_{M}}<\Omega_{\Lambda_{0}}< \Omega_{\Lambda_{c}}$.
For the plotting we put $\Omega_{M_{0}}=0.2$ and
$\Omega_{\Lambda_{0}}=0.4$. \label{fig9}}
\end{figure}

The present technology allows us to ``see" the epoch of last
scattering of radiation at a redshift of about $1000$, i.e., we
can actually observe the most distant information that the
Universe provides. The high level of isotropy observed from the
Cosmic Microwave Background Radiation (CMBR) \cite{Coles} from
this epoch to our days imposes that the two Hubble factors $H_a$
and $H_b$ must remain approximately equal from this epoch to the
present. In other words, we must impose a high isotropy level
from the last scattering onwards, in our expressions, i.e.,
$${{\Delta H} \over {H_a}}\equiv {{H_a-H_b} \over {H_a}},$$ such that
$\left| {{{\Delta H} \over {H_a}}} \right| \ll 1$. From COBE
4-Year data \cite{Bunn}, \cite{Kogut}, we have $\left( {{\sigma
\over H}} \right)_0 \sim 10^{-9}$ and for last scattering epoch
$\left( {{\sigma  \over H}} \right)_{ls} \sim 10^{-6}$. At the
last scattering we may still consider $H_a \simeq H_b \simeq H$
($H$ defined above).

We computed several numerical integrations, with equation (\ref{dif}),
by the following way: we gave values to $\Omega_{M_0}$ and
$\Omega_{\Lambda_0}$ and integrated back in time, from now to the last
scattering epoch. These $\Omega_{M_0}$ and $\Omega_{\Lambda_0}$ values
were chosen such that, at last scattering epoch we had
$\left| {{{\Delta H} \over H}} \right|_{ls} \equiv
\left|1- \left (\frac{dy}{dx} \right)_{ls}\right| = 1.7 \times 10^{-6}$ or
$\left (\frac{dx}{dy} \right)_{ls} = 1\pm 1.7 \times 10^{-6}$.
To make this we implemented a 8 order Runge-Kutta method
\cite{Hairer}.

We concluded that the sum $\Omega_{M_{0}}+\Omega_{\Lambda_{0}}$
must be close to the unity from above for Kantowski-Sachs and
from below for Bianchi type-III models\footnote{It is obvious that
for Bianchi type-I model
($\Omega_{M_{0}}+\Omega_{\Lambda_{0}}=1$), with our restrictions,
we have always $\Delta H/H_a =0$.}. We summarized in the table
below the result of imposing $\left| {{\Delta H  \over H}}
\right|_{ls} \sim 1.7 \times 10^{-6}$ for Kantowski-Sachs and
Bianchi type-III models, supposing $H_{a_{0}}=H_{b_{0}}$ (because
$\left( {{\sigma  \over H}} \right)_0 \sim 10^{-9}$).

\begin{table}
\caption{Density parameters and relative difference between $H_a$
and $H_b$ for Kantowsky-Sachs and Bianchi type-III models.
\label{tbl}} {\small
\begin{tabular}{cccccc}
\hline
& {$\Omega_{M_{0}}$}
& {$\Omega_{\Lambda_{0}}$}
& {$\Omega_0+\Omega_{\Lambda_{0}}$}
& {$\Omega_{k_{0}}$}
& {${{\Delta H} \over {H_a}}$} \\
\hline
\hline
K-S & $1$ & $\lesssim 1.7\times 10^{-8}$ & $1+5.6\times10^{-9}$ &
$-1.7\times 10^{-8}$ & $-1.6 \times 10^{-6}$ \\
K-S & $\lesssim 2\times10^{-15}$ & $1$ & $1+6.7\times10^{-16}$ &
$-2.0\times10^{-15}$ & $-1.4\times 10^{-6}$ \\
K-S & $\lesssim 0.3+7.0\times10^{-9}$ & $0.7$ & $1+2.3\times10^{-9}$ &
$-7.0\times10^{-9}$ & $-1.7\times 10^{-6}$ \\
K-S & $0.3$ & $\lesssim0.6+7.0\times10^{-9}$ & $1+2.3\times10^{-9}$ &
$-7.0\times10^{-9}$ & $-1.7\times 10^{-6}$ \\
B III & $1-10^{-10}$ & $\gtrsim9.9\times10^{-11}$ & $1-3.3\times10^{-13}$ &
$+1.0\times10^{-12}$ & $+1.3\times 10^{-6}$ \\
B III & $\gtrsim9.8\times10^{-14}$ & $1-10^{-13}$ & $1-6.7\times10^{-16}$ &
$+2.0\times10^{-15}$ & $+1.3\times 10^{-6}$ \\
B III & $\gtrsim0.3-10^{-11}$ & $0.7$ & $1-3.3\times10^{-12}$ &
$+1.0\times10^{-11}$ & $+1.8\times 10^{-6}$ \\
B III & $0.3$ & $\gtrsim 0.7-10^{-11}$ & $1-3.3\times10^{-12}$ &
$+1.0\times10^{-11}$ & $+1.8\times 10^{-6}$ \\
\hline
\end{tabular}}
\end{table}

From the table above we concluded that all combinations of
$\Omega_0+\Omega_{\Lambda_{0}}$ near the unity are equally
acceptable for reproducing a small anisotropy
($(\sigma/H)_{ls}\sim 10^{-6}$) at the last scattering.
Nevertheless we paid special attention to the values of
$\Omega_0\sim 0.3$ and $\Omega_{\Lambda_{0}}\sim 0.7$, since they
reproduce the better fit to recent observations \cite{Turner}. We
have in this scenario $|\Delta H/H|<2\times 10^{-6}$ for
Kantowski-Sachs and Bianchi type-III universes. All these models
approach locally an exponential expanding state \cite{Moniz}
provided the cosmological constant if we consider
$\Omega_\Lambda>\Omega_{\Lambda_{M}}$.

\section{Conclusions}

For the Kantowski-Sachs model ($\Omega_{k_0}<0$) (see Figures 4
and 5), we conclude that if the scale factor $b(t)$ starts from
zero, then the scale factor $a(t)$ will start from infinity and
decreases afterwards. When
$\Omega_{\Lambda_{0}}<\Omega_{\Lambda_{M}}$, $b(t)$ reaches the
maximum value recollapsing after that. So, $a(t)$ will reach a
relative maximum, when $b(t)$ is maximum, (see Figure 4). After
that, when $b(t)=0$, $a(t)$ goes to infinity again. When
$\Omega_{\Lambda_{M}}<\Omega_{\Lambda_{0}}<
\Omega_{\Lambda_{c}}$, the scale factor $b(t)$ grows indefinitely,
giving place to an inflationary scenario. Then, $a(t)$ decreases
reaching a minimum value, and growing after that indefinitely,
and becoming proportional to $b(t)$ (see Figure 5). The initial
singularity is of a ``cigar" type.

For the Bianchi type-I model ($\Omega_{k_{0}}=0$) (see Figures 6
and 7), the scale factors $a(t)$ and $b(t)$ are proportional or
even equal. Thus, this model turns out to be an isotropic one
(owing to our restrictions) and $\Omega_0+\Omega_{\Lambda_0}=1$.
However, when $\Omega_{\Lambda_{0}}< \Omega_{\Lambda_{M}}$,
$a(t)$ and $b(t)$ reach the maximum and recollapse after that.
And when $\Omega_{\Lambda_{M}}<\Omega_{\Lambda_{0}}<
\Omega_{\Lambda_{c}}$, $a(t)$ and $b(t)$ grow indefinitely after
an inflection.

For the Bianchi type-III model ($\Omega_{k_{0}}>0$) (see Figures 8
and 9), when $\Omega_{\Lambda_{0}}<\Omega_{\Lambda_{M}}$, $b(t)$
starts from an initial non vanishing value $(b(t=0)=b_0>0)$,
reaching a maximum and recollapsing after that until reaches the
same value for $t=0$. Also, $a(t)$ has a similar behavior, but
starts from zero and recollapses to zero,nevertheless, $a(t)$
exhibits a relative minimum when $b(t)$ is maximum. When
$\Omega_{\Lambda_{M}}<
\Omega_{\Lambda_{0}}<\Omega_{\Lambda_{c}}$, $b(t)$ starts again
from a non vanishing value $(b_0>0)$, growing indefinitely with
an inflection. In this case, $a(t)$ starts from zero and grows
indefinitely becoming approximately proportional to $b(t)$. So,
the initial singularity is of a ``pancake" type.

In conclusion, these models undergo isotropization becoming an
asymptotically FLRW, except for the Kantowski-Sachs model
$(\Omega_{k_{0}}<0)$ with $\Omega_{\Lambda_{0}}<
\Omega_{\Lambda_{M}}$ and for the Bianchi type-III
$(\Omega_{k_{0}}>)$ with
$\Omega_{\Lambda_{0}}<\Omega_{\Lambda_{M_{0}}}$. Taking into
account the accuracy of the measurements of anisotropy on one
hand and the fact that we can always adjust the density parameters
such that $|\Omega_0+\Omega_{\Lambda_0}-1|=\delta$, with $\delta
\sim 10^{-9}$ on the other, we conclude that these models can
still stand as good candidates to describe the observed Universe.
\vspace*{1cm}

{\noindent \bf \Large Acknowledgments}
\vspace*{.5cm}

The authors thank Alfredo B. Henriques, Jos\'e P. Mimoso
and Paulo Moniz for useful discussions and comments.
This work was supported in part
by grants BD 971 and BD/11454/97 PRAXIS XXI, from JNICT
and by CERN/P/FAE/1164/97 Project.

\end{document}